# Field-induced metal-to-insulator transition and colossal anisotropic magnetoresistance in a nearly Dirac material EuMnSb$_2$

Z. L. Sun[1*], A. F. Wang[2*], H. M. Mu[1], H. H. Wang[1], Z. F. Wang[1], T. Wu[1], Z. Y. Wang[1], X. Y. Zhou[2#] and X. H. Chen[1,3,4,5#]

[1]Hefei National Laboratory for Physical Sciences at Microscale and Department of Physics, and Key Laboratory of Strongly-Couple Quantum Matter Physics, Chinese Academy of Sciences，University of Science and Technology of China, Hefei, Anhui 230026, China.

[2]Center for Quantum Materials and Devices and Chongqing Key Laboratory of Soft Condensed Matter Physics and Smart Materials, College of Physics, Chongqing University, Chongqing 401331, China.

[3]CAS Center for Excellence in Superconducting Electronics (CENSE), Shanghai 200050, China.

[4]CAS Center for Excellence in Quantum Information and Quantum Physics, Hefei, Anhui 230026, China.

[5]Collaborative Innovation Center of Advanced Microstructures, Nanjing University, Nanjing 210093, China.

**ABSTRCT**

**How to realize applicably appreciated functionalities based on the coupling between charge and spin degrees of freedom is still a challenge in the field of spintronics. For example, anisotropic magnetoresistance (AMR) effect is utilized to read out the information stored by various magnetic structures, which usually originates from atomic spin-orbit coupling (SOC). However, the application of AMR in antiferromagnet-based spintronics is still hindered by rather small AMR value. Here, we discover a colossal AMR effect during the field-induced metal-to-insulator**

**transition (MIT) in a nearly Dirac material EuMnSb$_2$ with an antiferromagnetic order of Eu$^{2+}$ moments. The colossal AMR reaches to an unprecedented value of $1.84 \times 10^6$% at 2 K, which is four orders of magnitude larger than previously reported values in antiferromagnets. Based on density functional theory calculations, a Dirac-like band structure, which is strongly dependent on SOC, is confirmed around *Y* point and dominates the overall transport properties in the present sample with predominant electron-type carriers. Moreover, it is also revealed that the indirect band gap around Fermi level is dependent on the magnetic structure of Eu$^{2+}$ moments, which leads to the field-induced MIT and plays a key role on the colossal AMR effect. Finally, our present work suggests that the similar antiferromagnetic topological materials as EuMnSb$_2$, in which Dirac-like fermions is strongly modulated by SOC and antiferromagnetism, would be a fertile ground to explore applicably appreciated AMR effect.**

## Ⅰ. INTRODUCTION

The manipulation of charge transport by spin degree of freedom in solid-state systems is at the core of spintronics. Recently, antiferromagnets have generated significant interest in the field of spintronics owing to their unique properties, such as zero stray magnetic field, ultrafast spin dynamics, and remarkable rigidity against external fields [1-4]. Anisotropic magnetoresistance (AMR), which is defined as the dependence of the resistivity on the direction of the magnetization with respect to current or crystal axes, could be utilized as electrical-readout to detect the switch of magnetization in antiferromagnetic (AFM)-based spintronic devices. AMR-based memory effects have been successfully demonstrated in several antiferromagnets, such as FeRh [5] and MnTe [6], opening perspective for both fundamental research and device technology of AFM spintronics [5-11]. In principle, the conventional AMR effect, in which the electronic band structure is dependent on the spin orientation, is mainly associated with the magnetocrystalline anisotropy arising from the relativistic spin-orbit coupling (SOC). In this sense, the SOC-induced anisotropic response of band structure to external magnetic field and the detailed position of the Fermi level determine the AMR in practical antiferromagnets [12-15]. However, experimentally observed AMR is always limited to a few hundred percent, which is much smaller than the giant magnetoresistance (GMR) [16,17] or tunneling magnetoresistance (TMR), hindering its further practical application. Thus, the hunt for new materials with large AMR and new strategies toward enhancement of AMR are highly desired.

On the other hand, the rise of the magnetic topological materials provides a new fertile playground to explore the manipulation of quantum transport phenomena by spin degrees of freedom [2]. In recent years, the strong correlation between magnetism and topological band structure has been widely discussed in magnetic topological materials [18,19]. Especially, the topological band structure can be strongly dependent on the strength of SOC and the detailed magnetic structure in some magnetic topological materials [20-22], which leads to anisotropic response of band structure near the Fermi level to the change of magnetic structure. Therefore, the magnetic topological materials can potentially generate a large AMR, which has not yet been observed experimentally. Among various magnetic topological materials, the family of $A$Mn$Pn_2$ ($A$ = Ca, Sr, Ba, Eu or Yb, $Pn$ = Sb or Bi) materials, in which the conducting two-dimensional (2D) Bi/Sb layers separated by the insulating Mn-$Pn$ layers with magnetism can host Dirac-like fermions, exhibits great potential for strong correlation between the Dirac-like band structure and magnetism. In principle, the 2D Bi/Sb layers with square net structure can host 2D Dirac-like fermions [23,24], and quasi-2D massive Dirac-like fermions with SOC gap have been observed in $A$Mn$Pn_2$ with Bi/Sb square net [23,25-31], including SrMnBi$_2$ [23,25] and CaMnBi$_2$ [26]. On the other hand, although theoretical study [32] has argued that the formation of zigzag chain structure in 2D Bi/Sb layers will destroy the linear band dispersion held by the square net structure, the existence of non-trivial Dirac fermions have also been verified in the *A*Mn*Pn*$_2$ family materials with zigzag chain structure by quantum oscillation experiments, including SrMnSb$_2$ [33], CaMnSb$_2$ [34] and BaMnSb$_2$ [35,36]. Angle resolved photoemission spectroscopy (ARPES) experiments have further confirmed a linearly Dirac-like dispersive band structure around Fermi level in EuMnSb$_2$ and BaMnSb$_2$ [35-37]. These results strongly indicate that holding a Dirac-like dispersion seems to be a generic nature of band structure in the *A*Mn*Pn*$_2$ family materials. More interestingly, a considerable coupling between the magnetism and the Dirac-like fermions in the conducting Bi/Sb layer has been clearly observed by neutron scattering and Raman-scattering measurement [38-41] in various *A*Mn*Pn*$_2$ family materials. Especially, when *A* represents magnetic ion Eu$^{2+}$, the interaction between local moments and itinerant carrier could be significantly enhanced due to the highly tunable magnetic moments of Eu$^{2+}$ by external magnetic field. For instance, quantum Hall effect is observed in bulk antiferromagnet EuMnBi$_2$, in which the field-controlled Eu$^{2+}$ magnetic order suppresses the interlayer coupling [28]. In contrast to the positive magnetoresistance (MR) observed in other members of this family, EuMnSb$_2$ with zigzag chain structure in 2D Sb layers shows a large negative MR, especially for the temperature range below the AFM ordering temperature of Eu$^{2+}$ moments [37,42,43]. All these results strongly suggest a strongly modulated Dirac-like band structure by the magnetism of Eu$^{2+}$ moment in the EuMn*Pn*$_2$ materials.

Here, we discover a colossal AMR effect during the field-induced metal-to-insulator transition (MIT) in a nearly Dirac material EuMnSb$_2$ with an antiferromagnetic order of Eu$^{2+}$ moment. The colossal AMR reaches to an unprecedented value of 1.84×10$^6$% at 2 K, which is four orders of magnitude larger than previously reported values in antiferromagnets. Our density functional theory (DFT) calculations indicates that the field-induced MIT and colossal AMR effect are related to a Dirac-like band structure around $Y$ point, which is strongly modulated by SOC and antiferromagnetism.

## Ⅱ. METHOD

### A. Crystal growth and characterization

EuMnSb$_2$ single crystals used in this study were grown by using Sn flux method. A mixture of Eu slugs, Mn pieces, Sb slugs and Sn shots in a molar ratio of 1:1:2:10 was loaded in an alumina crucible and sealed in an evacuated quartz tube under vacuum. The quartz tube was heated slowly to 1000℃ and held for 20h to homogenize the melt. Then it was cooled down to 600℃ at a rate of 2℃/h, where the flux was decanted using a centrifuge. The compositions of crystals were determined using an energy-dispersive X-ray spectrometer (EDS) mounted on the field emission scanning electronic microscope (FESEM), Sirion200. The single crystal X-ray diffraction (XRD) pattern was obtained by the X-ray diffractometer (SmartLab-9, Rigaku Corp.) with Cu Kα radiation and a fixed graphite monochromator.

### B. Electrical transport and magnetic measurements

Electric transport measurements were carried out by using *Quantum Design* PPMS-14 equipped with a rotator module. Resistivity was measured using standard four-probe method and Hall resistivity was measured by standard Hall bar configuration. The rotation measurement was performed by using a commercial rotator in PPMS-14. Magnetization measurements were performed using *Quantum Design* MPMS-7.

## C. Density functional calculations

Spin-polarized DFT+U calculations were performed within the Vienna ab initio Simulation Package (VASP) [44]. The generalized gradient approximation (GGA) [45] and the Perdew-Burke-Ernzerh (PBE) [45,46] of exchange-correlation functional were used. The electron-ion interaction was described with the projector augmented wave (PAW) method and Eu (4f, 5s, 5p, 6s), Mn (3d, 4s), Sb (5s, 5p) electrons were treated as valence states. An effective Hubbard U = 5 eV in the GGA+U scheme [47] was applied to the localized f (Eu) and d (Mn) electrons, and the spin-orbit coupling interactions were also taken into account in the DFT calculations.

## III. RESULTS

### A. Magnetic and transport properties

$EuMnSb_2$ crystallizes in an orthorhombic structure with space group *Pnma* (No.62) as shown Fig. 1(d). The single crystal XRD pattern shows high quality of our sample with lattice parameter *a* = 22.56 Å (shown in Fig. S1 of the Supplementary Materials [48]), which is consistent with the previous report [37,42,43]. The Mn sublattice, sandwiched by two Sb atomic layers, exhibits C-type AFM order with magnetic moments aligned along the *a* axis below ~346 K [37,43]. So far, there is a discrepancy between the reported magnetic structures of Eu sublattice below 21 K. A-type AFM structure with magnetic moments along the *c* axis was suggested by powder neutron diffraction measurements [37], while neutron diffraction refinements performed on single crystals have found a canted A-type AFM order with $Eu^{2+}$ moments lying 41 ° away from the *a* axis in the *ac* plane [43]. To confirm the magnetic order of Eu sublattice in our $EuMnSb_2$ single crystal, we have performed magnetization measurements.

The temperature-dependent magnetic susceptibilities $\chi_a$ and $\chi_{bc}$ are displayed in Fig. 1(a), with magnetic field ($H$) applied along the $a$ axis and $bc$ plane, respectively. $\chi_a$ shows a clear kink at 21 K, denoting the formation of AFM order of Eu sublattice. In Fig. 1(b), we display the isothermal magnetization of EuMnSb$_2$ with magnetic field applied along the $a$ axis and $bc$ plane at 2 K, respectively. In both cases, the magnetization does not saturate up to 7 T. When $H$ is applied along the $a$ axis, a spin-flop transition is observed at $H$ = 1.5 T (red curve), consistent with previous report [42]. The magnetization along the $a$ axis is slightly smaller than that in the $bc$ plane, indicating that the magnetic moments of Eu$^{2+}$ are more easily polarized in the $bc$ plane. The small anisotropy of magnetization between the $a$ axis and $bc$ plane implies a preferred canted A-type AFM structure of Eu$^{2+}$ moments in EuMnSb$_2$ as shown in Fig. 1(d). As we will discuss later, our DFT calculations also find that the total energy with canted A-type AFM structure of Eu$^{2+}$ moments is smaller than that with A-type AFM structure. Therefore, we consider the canted A-type AFM order as the magnetic ground state of Eu sublattice in the relevant discussion of the present manuscript. More magnetic susceptibility data can be found in Fig. S2 of the Supplementary Materials [48].

Fig. 1(c) shows temperature dependence of in-plane resistivity $\rho_{xx}$ for EuMnSb$_2$ with external magnetic field applied along the $a$ axis. At zero field, it shows a semiconducting behavior with a kink appearing at the AFM transition temperature of the Eu sublattice ($T_N$ ~ 21 K). The longitudinal $\rho_{xx}$ at zero magnetic field can be fitted by the Arrhenius equation $\rho_{xx}(T) \propto \exp(E_A/2k_BT)$ from 200 K to 300 K, where $E_A$ is the thermal activation energy and $k_B$ is the Boltzmann constant (shown in Fig. S3 of the Supplementary Materials [48]). The fitting procedure gives an activation energy of 75.9 meV which is larger than that in previous report [42]. The deviation of resistivity below 200 K from thermally activated resistivity behavior can be attributed to the short-range magnetic order of Eu$^{2+}$ moments, which is also consistent with previous report [43]. We note that the resistivity of our EuMnSb$_2$ sample at zero magnetic field shows insulating behavior below $T_N$, which is quite different from that in previous report exhibiting a dip below $T_N$ [37,42,43]. In addition, the value of resistivity in our EuMnSb$_2$ sample is also much larger than that in previously reported sample [37,42,43]. We ascribe the insulating behavior of resistivity to a lower carrier density in our EuMnSb$_2$ sample, thus making the Fermi level locate inside the band gap. This will be discussed below in detail. When $H$ is applied along the $a$ axis, it induces a MIT at low temperature in our EuMnSb$_2$ sample. With increasing magnetic field, the small kink in resistivity corresponding to $T_N$ moves to lower temperature and the upturn behavior below $T_N$ becomes weak with increasing magnetic field. This indicates that the band gap below $T_N$ decreases under magnetic field, which is consistent with the following band structure calculations. The resistivity $\rho_{xx}$ with the magnetic field perpendicular and parallel to the current direction in the $bc$ plane is shown in Fig. 1(e) and 1(f), respectively. $\rho_{xx}$ with $H//bc$ shows relatively weak negative MR behavior and remains insulating behavior at low temperature even at 14 T, which is quite different from the nearly metallic behavior with $H//a$.

The magnetic-field-dependent resistivity data shown in Fig. 1(c) reveals that EuMnSb$_2$ exhibits a remarkable negative MR response at low temperatures. To proceed, we define MR as MR = $[\rho_{xx}(H) - \rho_{xx}(0\,\text{T})]/\rho_{xx}(0\,\text{T})$ and display the MR at different temperatures with magnetic field applied along the $a$ axis in Fig. 1(g). The MR has been symmetrized with respect to positive and negative magnetic fields to remove the Hall contribution. EuMnSb$_2$ shows a large negative MR below $T_N$ and above about 5 T. The negative MR behavior weakens with temperature increasing above $T_N$. A large negative MR with maximum value of -99.9987% at 2 K and under $H$= 14 T is observed, which is larger than the value in previously reported EuMnSb$_2$ by far [37,42]. To gain more sight into the emergence of the negative MR, we have plotted the temperature-dependent MR at various magnetic fields applied along the *a* axis in Fig. 1(h). The large negative MR response sets in when the magnetic moments of $Eu^{2+}$ form AFM order, and disappears quickly above $T_N$, indicating a clear correspondence between the negative MR and magnetic order of Eu sublattice. Previously, negative MR has been observed in some low-carrier density magnetic Eu-based materials [49-52]. The negative MR was ascribed to the formation of magnetic polaron, for example, in $Eu_5In_2Sb_6$ [51] and $EuB_6$ [52]. Nevertheless, ac magnetic susceptibility measurement in $EuMnSb_2$ [42] does not support the existence of magnetic polaron, which excludes the magnetic polaron as possible origin of the negative MR. Instead, the germane correlation between the charge-transport and magnetization implies that the large negative MR observed in $EuMnSb_2$ is related to the magnetic order of $Eu^{2+}$ moments, which can be further demonstrated by the following band structure calculations.

**B. Anisotropic magnetoresistance**

Conspicuously, the MR also shows extremely anisotropic when the magnetic field rotates from the *a* axis, as shown in Fig. 2(a). The set-up geometry of our measurements is illustrated in the inset of Fig. 2(a). Here the electric current *I* is applied in the *bc* plane while magnetic field *H* is rotated from the *a* axis by angle $\theta$ with keeping perpendicular to *I*. All curves with varied angles show a negative MR, and the resistivity at high field increases drastically with $\theta$ increasing. Fig. 2(b) displays the AMR = $[\rho_{xx}(\theta)-\rho_{xx}(0°)]/\rho_{xx}(0°)$ [53,54] as a function of $\theta$ at 2 K under several representative magnetic fields. It clearly shows a remarkably large AMR with maximum value $1.84\times10^6$% at 12 T, which exceeds previously reported AMR values in AFM materials by far. A complete AMR contour map with magnetic field larger than 5 T is presented in Fig. 2(c). The AMR shows two-fold symmetry and the center located at $H=$ 12 T and $\theta=$ 90 °. Fig. 2(d) displays the magnetic-field-dependent AMR at 90 °, and one can see that the AMR shows a relatively small value below 5 T, above which it increases rapidly and shows the maximum at 12 T. The AMR data at low magnetic fields are shown in Fig. S4 of the Supplementary Materials [48]. In fact, the relatively smaller AMR value at 5 T already reaches to ~800%.

To reveal the temperature dependence of the colossal AMR, we have measured the AMR at various temperatures under $H=$ 12 T, as depicted in Fig. 2(e). The two-fold AMR decreases monotonically with increasing temperature. Fig. 2(f) shows the temperature-dependent AMR value at $H=$ 12 T and $\theta=$ 90 °. The AMR rapidly decreases with increasing temperature to $T_N$, and shows a much smaller value above $T_N$. It further implies a close relation between the colossal AMR and the AFM order of $Eu^{2+}$ moments. When the magnetic field is rotated in the $bc$ plane, it shows relatively small AMR, as displayed in Fig. S5 of the Supplementary Materials [48].

**C. Electronic structure calculations**

To investigate the origin of the colossal AMR and the topological nature of our $EuMnSb_2$ sample, we have calculated the electronic band structure with different $Eu^{2+}$ magnetic orders. First, we perform the band structure calculation of $EuMnSb_2$ in the AFM state of Eu sublattice without SOC [Fig. 3(a)]. The band structure at $Y$ point shows a linear band dispersion with a tiny gap, indicating $EuMnSb_2$ is a massive Dirac semimetal without considering SOC, just similar to the other members in the $AMnPn_2$ family. Next, we performed band structure calculations considering canted A-type AFM and $c$-axis-oriented A-type AFM orders of Eu sublattice with SOC, as shown in Fig. 3(b) and 3(c), respectively. The resulted total energy in the former case is about 10 meV less than that in the latter, indicating that the Eu sublattice in EuMnSb$_2$ favors the canted A-type AFM order at zero magnetic field. As mentioned earlier, this finding is consistent with our magnetic susceptibility measurements.

Our band structure calculation reveals a strong SOC effect at *Y* point. Due to the effect of SOC, the band structure at *Y* point opens a sizable gap with 162 meV for canted A-type AFM order of Eu sublattice and 146 meV for *c*-axis-oriented A-type AFM order of Eu sublattice. Despite the existence of band gap, the band structure shows a highly dispersive and linear band at *Y* point, manifesting EuMnSb$_2$ is a nearly Dirac semimetal with a gapped Dirac point. The states near the Fermi level mainly comes from the 2D Sb zigzag layer (as shown in Fig. S6 of the Supplementary Materials [48]), which indicates the charge carriers in Sb layers are Dirac-like. We can understand the generation of the gapped Dirac point at *Y* point from the orthorhombic distortion from the Sb tetragonal square. For tetragonal *A*MN*Pn*$_2$ with Sb/Bi square net, the Dirac-like band exists along the along *Γ-M* line [23,24,55]. The orthorhombic zigzag distortion in EuMnSb$_2$ opens a sizable gap ~0.5 eV at this pristine Dirac point on the *Γ-M* line and generates a highly dispersive linear band at *Y* point. The band structure of EuMnSb$_2$ is similar to the result in BaMnSb$_2$ with Sb zigzag chain. Considering a strong SOC, the band structure of BaMnSb$_2$ with Sb zigzag chains could still show a Dirac-like linear dispersion around *Y* point with a small band gap, which is supported by the quantum oscillation experiment and ARPES measurement [35,36]. Furthermore, the nontrivial Berry phase in quantum oscillation experiment has also been revealed in similar family materials with Sb zigzag chains, including SrMnSb$_2$ [33] and CaMnSb$_2$ [34], which have the same space group (*Pnma*) as EuMnSb$_2$. Therefore, a gapped Dirac cone around *Y* point seems to be a generic topological nature in these Mn-based 112 materials with Sb zigzag chains. Consequently, we think that EuMnSb$_2$ could be a nearly Dirac semimetal with gapped Dirac fermions. In addition, the calculations show an indirect band gaps between the bottom of conductive band at *Y* point and the top of the valence band in the *Γ-M* line with 148.9 meV for the canted A-type AFM order and 125.3 meV for *c*-axis-oriented A-type AFM order.

Then, we also calculate the band structures with ferromagnetic (FM) order of Eu sublattice. Since the Mn sublattice forms robust C-type AFM order at much higher temperature (about 346 K) [37,43], we consider that the magnetic order of Mn moments remains unchanged at low temperatures with the magnetic field we have applied. In fact, magnetization measurements show that the magnetic order of Mn sublattice does not change up to 55 T in a related compound, EuMnBi$_2$ [28], indicating that Mn moments are extremely hard to be modified in this series of materials. Therefore, the magnetic structure of Mn sublattice is fixed to C-type AFM in our band structure calculations. As shown in Fig. 3(d), the degenerate band in the FM state of Eu sublattice splits into spin-up and spin-down channel without SOC. The FM order has strong effect on the band around the *Γ* point near the Fermi level, resulting in a large splitting, while it has weak effect on the Dirac-like band at *Y* point. The calculated band structure considering SOC with Eu$^{2+}$ moments ferromagnetically aligned along the *a* and *c* axis are displayed in Fig. 3(e) and 3(f), respectively. Remarkably, a strong modulation of the band structure with the reorientation of Eu$^{2+}$ moments is found: When the Eu$^{2+}$ moments are aligned along the *a* axis, the indirect band gap reduces to 23.3 meV; when the Eu$^{2+}$ moments are oriented in the *c* axis, the indirect band gap is found to be 62.0 meV, which is a bit larger than the value with Eu moments along the *a* axis. The change of band structure by the spin orientation of Eu$^{2+}$ moments in EuMnSb$_2$ leads to significant magneto-transport response. At zero field, EuMnSb$_2$ shows a relatively large band gap ~ 148.9 meV, resulting in semiconducting behavior of the resistivity below $T_N$. When the magnetic field is applied along the *a* axis or *c* axis, the Eu$^{2+}$ moments orient toward the field direction progressively, giving rise to a large reduction of the band gap. This leads to the negative MR observed below $T_N$.

Owing to the exchange interaction between $Eu^{2+}$ and $Mn^{2+}$ moments, a large magnetic field is needed to fully polarize the $Eu^{2+}$ moments (e.g., about 22 T in $EuMnBi_2$ [28]). Therefore, the MR does not saturate up to 14 T as shown in Fig. 1(g) and 2(a), and should further increase until the $Eu^{2+}$ moments are fully polarized. To further establish the relationship between the band gap and magnetic structure, we have calculated the band gap values based on various magnetic orders of Eu sublattice with magnetic moments oriented $\theta$ away from the $c$ axis in the $ac$ plane, as shown in Fig. S7 of the Supplementary Materials [48]. It is clear that the band gap continuously decreases as $Eu^{2+}$ moments rotating from the $c$ axis to $a$ axis. When the magnetic field applied along the $a$ axis increases, the $Eu^{2+}$ moments trend to gradually polarize along the $a$ axis (increasing $\theta$), giving rise to a reduction of the band gap. As all the $Eu^{2+}$ moments are oriented along the $a$ axis ferromagnetically ($\theta = 90°$), the smallest band gap with a value ~23.3 meV is achieved. We note that while the heuristic calculation considers A-type AFM order with $Eu^{2+}$ moments aligned along the $c$ axis as the starting point, which differs from the canted A-type AFM order in real case, it serves to motivate the general evolution of band gap against external magnetic field in $EuMnSb_2$. In addition, the fitted excitation gap with $E_A$ = 75.9 meV above 200 K is smaller than the calculated band gap of 148.9 meV in canted AFM state but larger than that in field-induced AF state. This indicates that the short-range magnetic correlation in high-temperature paramagnetic state above $T_N$ can also give rise to a reduction in the band gap, which results in the transport behavior above $T_N$.

## Ⅳ. DISCUSSION

We further analyze the origin of the colossal AMR in our $EuMnSb_2$ sample. The band structure calculations find anisotropic band gaps between different spin orientations: 23.3 meV with $Eu^{2+}$ moments aligned along the *a* axis and 62.0 meV with $Eu^{2+}$ moments aligned along the *c* axis, respectively. The rotating magnetic field from the *a* axis to the *c* axis enforces the orientation of the $Eu^{2+}$ moments, resulting in different band gap values. Such anisotropic band gap combined with corresponding position of Fermi level can result in the colossal AMR. The detailed evolution of the band structure and the position of Fermi level under magnetic field is illustrated in Fig. 4. When the magnetic field is below 5 T, the Fermi level remains inside the band gap for both *H*//*a* and *H*//*c*, which leads to a moderate negative MR effect [Fig. 2(a)] and almost negligible AMR effect [Fig. 2(d)]. Due to the Fermi level inside the band gap, the spin-flop transition around 1.5 T (as shown in Fig. 1(b)) has no significant effect on both MR and AMR. As the applied external field above 5 T, a field-induced MIT occurs and the Fermi level starts to cross the conduction band around *Y* point with *H*//*a* but still remains inside the band gap with *H*//*c*. Then, a significant MR effect only appears with *H*//*a*, but not with *H*//*c* (see a fast drop of resistivity with *H*//*a* in Fig. 2(a)), which leads to a colossal AMR effect. When the applied external field further increases above 12 T, the Fermi level is also across the conduction band around *Y* point with *H*//*c*. Then, a significant MR effect also appears with *H*//*c*, which leads to the decreasing of AMR effect above 12 T [Fig. 2(d)]. Therefore, we think that there are two key factors to account for above colossal AMR effect. One is the magnetic-structure-induced anisotropic reduction of band gap. The other is that the position of Fermi level should be across the conduction band during the anisotropic reduction of band gap by applying magnetic field, which actually leads to a characteristic behavior for MIT in resistivity.

The second factor mentioned above also offers a natural explanation on the different behavior between our sample and the samples used in previous report. The zero-field resistivity of our $EuMnSb_2$ sample shows a different low-temperature behavior from that in previous reports, as shown in Fig. S8 of the Supplementary Materials [48]. In previous reported EuMnSb$_2$ grown by flux or floating-zone method [37,42,43], the low-temperature resistivity shows a metallic or weak insulating behavior. However, in our EuMnSb$_2$ single crystal, there is a strong insulating behavior at low temperature, especially below $T_N$. Then, the value of resistivity at 2K in our samples becomes almost three order of magnitude larger than that in literatures. Our EuMnSb$_2$ sample has much lower carrier density $9.67\times10^{17}$ cm$^{-3}$ (as shown in Fig. S9 of the Supplementary Materials [48]) than the value $\sim3.8\times10^{18}$ cm$^{-3}$ in previous report [42]. Consequently, we attribute such difference in resistivity to the unintentional carrier doping induced by the impurity from Sn flux or Mn vacancy in previously reported samples. The low doping level makes the Fermi level locate in the band gap in our sample and the measured temperature-dependent resistivity behaves like an intrinsic antiferromagnetic semiconductor below $T_N$. However, the Fermi level lies near or intersect the band edge due to the higher carrier density in previously reported EuMnSb$_2$ [37,42,43], thus causing weakly insulating resistivity behavior with a drop or upturn behavior at low temperature. In despite of the existence of anisotropic band gap under magnetic field, the Fermi level across the band results in a much smaller MR and AMR values in previously reported EuMnSb$_2$ samples [37,42,43]. In addition, our Hall resistivity indicates that the carriers in our EuMnSb$_2$ are electron-type [48], which is different from the hole carrier in previous report [42]. The different carrier type also indicates that the AMR effect comes from different mechanism between our EuMnSb$_2$ sample and previously reported samples. The dominant electron-type carrier in our EuMnSb$_2$ sample coming from the $Y$ point is strongly affected by the SOC. Thus, our EuMnSb$_2$ can show colossal AMR effect due to the strong SOC effect on the electron-type carrier. However, the hole-type carrier in previously reported material [42] mainly comes from the top of the valance band along the $\varGamma$-$M$ line, which has a weak SOC effect, but is strongly affected by the FM order of Eu sublattice. Consequently, previously reported EuMnSb$_2$ is weakly affected by the SOC, thus showing a moderate AMR effect with relatively smaller value.

## Ⅴ. CONCLUSIONS

In conclusion, we have observed a colossal AMR reaching ~$1.84\times10^6$% at 2 K and field-induced MIT in a nearly Dirac semimetal EuMnSb$_2$ with an antiferromagnetic order of Eu$^{2+}$ moments. The external magnetic field can significantly modify the band structure at $Y$ point near the Fermi level, which is strongly dependent on SOC and magnetism. Our work establishes magnetic topological materials as an excellent candidate to realize strong response of magneto-transport properties to the magnetism. The revealed mechanism for colossal AMR in present study could be widely used to the topological materials in which Dirac-like fermions is strongly modulated by SOC and antiferromagnetism, which is important for the application of topological materials on spintronics.

**Acknowledgements**

We thank D. Z. Hou for the helpful discussion. X. H. Chen acknowledges Anhui Initiative in Quantum Information Technologies (AHY160000), the National Key Research and Development Program of the Ministry of Science and Technology of China (2016YFA0300201 and 2017YFA0303001), the Science Challenge Project of China (Grant No. TZ2016004), the Key Research Program of Frontier Sciences, CAS, China (QYZDYSSW-SLH021), the Strategic Priority Research Program of Chinese Academy of Sciences (XDB25000000), and the National Natural Science Foundation of China (11888101 and 11534010). A. F. Wang acknowledges National Natural Science Foundation of China (Grant No. 12004056) and Projects of President Foundation of Chongqing University

(2019CDXZWL002).

**Correspondence:** Correspondence and requests for materials should be addressed to X. H. Chen (email: chenxh@ustc.edu.cn) and X. Y. Zhou (email: xiaoyuan2013@cqu.edu.cn)

*Z. L. Sun and A. F. Wang contribute equally to this work.

Phys. Condens. Matter **26**, 042201 (2014).

33. J. Y. Liu, J. Hu, Q. Zhang, D. Graf, H. B. Cao, S. M. A. Radmanesh, D. J. Adams, Y. L. Zhu, G. F. Cheng, X. Liu, W. A. Phelan, J. Wei, M. Jaime, F. Balakirev, D. A. Tennant, J. F. DiTusa, I. Chiorescu, L. Spinu, and Z. Q. Mao, *A magnetic topological semimetal $Sr_{1-y}Mn_{1-z}Sb_2$ (y,z<0.1)*. Nat. Mater. **16**, 905 (2017).

34. J. B. He, Y. Fu, L. X. Zhao, H. Liang, D. Chen, Y. M. Leng, X. M. Wang, J. Li, S. Zhang, M. Q. Xue, C. H. Li, P. Zhang, Z. A. Ren, and G. F. Chen, *Quasi-two-dimensional massless Dirac fermions in $CaMnSb_2$*, Phys. Rev. B **95**, 045128 (2017).

35. H. Sakai, H. Fujimura, S. Sakuragi, M. Ochi, R. Kurihara, A. Miyake, M. Tokunaga, T. Kojima, D. Hashizume, T. Muro, K. Kuroda, T. Kondo, T. Kida, H. Hagiwara, K. Kuroki, M. Kondo, K. Tsuruda, H. Murakawa, and N. Hanansaki, *Bulk quantum Hall effect of spin-valley coupled Dirac fermions in the polar antiferromagnet $BaMnSb_2$*. Phys. Rev. B **101**, 081104(R) (2020).

36. J. Y. Liu, J. Yu, J. L. Ning, H. M. Yi, L. Miao, L. J. Min, Y. F. Zhao, W. Ning, K. A. Lopez, Y. L. Zhu, T. Pillsbury, Y. B. Zhang, Y. Wang, J. Hu, H. B. Cao, F. Balakirev, F. Weickert, M. Jaime, Y. Lai, Kun Yang, J. W. Sun, N. Alem, V. Gopalan, C. Z. Chang, N. Samarth, C. X. Liu, R. D. Mcdonald, Z. Q. Mao, *Spin-valley locking, bulk quantum Hall effect and chiral surface state in a noncentrosymmetric Dirac semimetal $BaMnSb_2$*. Preprint at https://arxiv.org/abs/1907.06318.

37. J. -R. Soh, P. Manuel, N. M. B. Schröter, C. J. Yi, F. Orlandi, Y. G. Shi, D. Prabhakaran, and A. T. Boothroyd, *Magnetic and electronic structure of Dirac semimetal candidate $EuMnSb_2$*, Phys. Rev. B **100**, 174406 (2019).

38. A. Sapkota, L. Classen, M. B. Stone, A. T. Savici, V. O. Garlea, A. Wang, J. M. Tranquada, C. Petrovic, and I. A. Zaliznyak, *Signatures of coupling between spin waves and Dirac fermions in*

**Figure captions**

FIG. 1. Resistivity, magnetoresistance and susceptibility of EuMnSb$_2$. (a) Temperature dependence of susceptibilities $\chi_a$ and $\chi_{bc}$ with magnetic field of 0.1 T applied along the *a* axis and *bc* plane. (b) Isothermal magnetizations of EuMnSb$_2$ with magnetic field paralleled to the *a* axis and *bc* plane at 2 K. (c) Temperature-dependent resistivity with magnetic fields applied along the *a* axis. (d) Schematic illustration of the magnetic structure of EuMnSb$_2$ at zero magnetic field. Eu sublattice forms canted A-type AFM order and Mn sublattice forms C-type AFM order. (e) $\rho_{xx}$ under different magnetic fields parallel to the current in the *bc* plane. (f) $\rho_{xx}$ under different magnetic fields perpendicular to the current in the *bc* plane. (g) MR (defined as MR = $[\rho_{xx}(H)-\rho_{xx}(0\ \text{T})]/\rho_{xx}(0\ \text{T})$) of EuMnSb$_2$ at varying temperatures with magnetic field along the *a* axis. (h) MR as a function of temperature with magnetic fields applied along the *a* axis.

FIG. 2. AMR of EuMnSb$_2$. (a) Resistivity with magnetic field rotating from the *a* axis to *bc* plane at 2 K. Inset: sketch of magneto-transport measurement geometries. The electric current is applied in *bc* plane and the magnetic field is rotated from the *a* axis to the basal plane with keeping perpendicular to the current. The angle $\theta$ is measured from the *a* axis to the magnetic field direction. (b) Angular-dependent AMR (defined as AMR = $[\rho_{xx}(\theta)-\rho_{xx}(0\ °)]/\rho_{xx}(0\ °)$) at 2 K under representative magnetic fields. Here, the direction that we define as 0 ° has an offset about 5 ° from the *a* axis, which probably originates from the misalignment between the crystal axis of sample with the rotator. (c) Color plot of AMR map at 2 K under magnetic field between 5 T and 14 T. (d) Magnetic field dependence of AMR value of 90 ° at 2 K. It increases rapidly with increasing magnetic field up to ~ 5 T, and shows a maximum at 12 T. (e) Evolution of AMR with the angle between the *a* axis and magnetic field of 12 T at various temperatures. (f) AMR value obtained at 90 ° as a function of temperature under 12 T. It shows relatively small value above the AFM transition temperature $T_N$ of Eu sublattice.

FIG. 3. Band structure of EuMnSb$_2$ with different orientations of Eu$^{2+}$ magnetic moments. (a) Band structure of EuMnSb$_2$ with Eu sublattice forming AFM order without SOC. (b) Left: Band structure of EuMnSb$_2$ with Eu sublattice forming canted A-type AFM order. Right: Schematic illustration of the canted A-type AFM order of Eu$^{2+}$ moments. (c) Left: Band structure of EuMnSb$_2$ with Eu sublattice forming A-type AFM order with moments along the *c* axis. Right: Schematic illustration of the of A-type AFM order of Eu$^{2+}$ moments. (d) Band structure of EuMnSb$_2$ with Eu sublattice forming FM order without SOC. (e) Left: Band structure of EuMnSb$_2$ with FM order of Eu$^{2+}$ moments aligned along the *a* axis. Right: Schematic illustration of the magnetic structure of Eu sublattice with FM order along *a* axis. (f) Left: Band structure of EuMnSb$_2$ with FM order of Eu$^{2+}$ moments aligned along the *c* axis . Right: Schematic illustration of the magnetic structure of Eu sublattice with FM order along the *c* axis.

FIG. 4 The evolution of band structure and corresponding Fermi level under magnetic field.

**Figure 1**

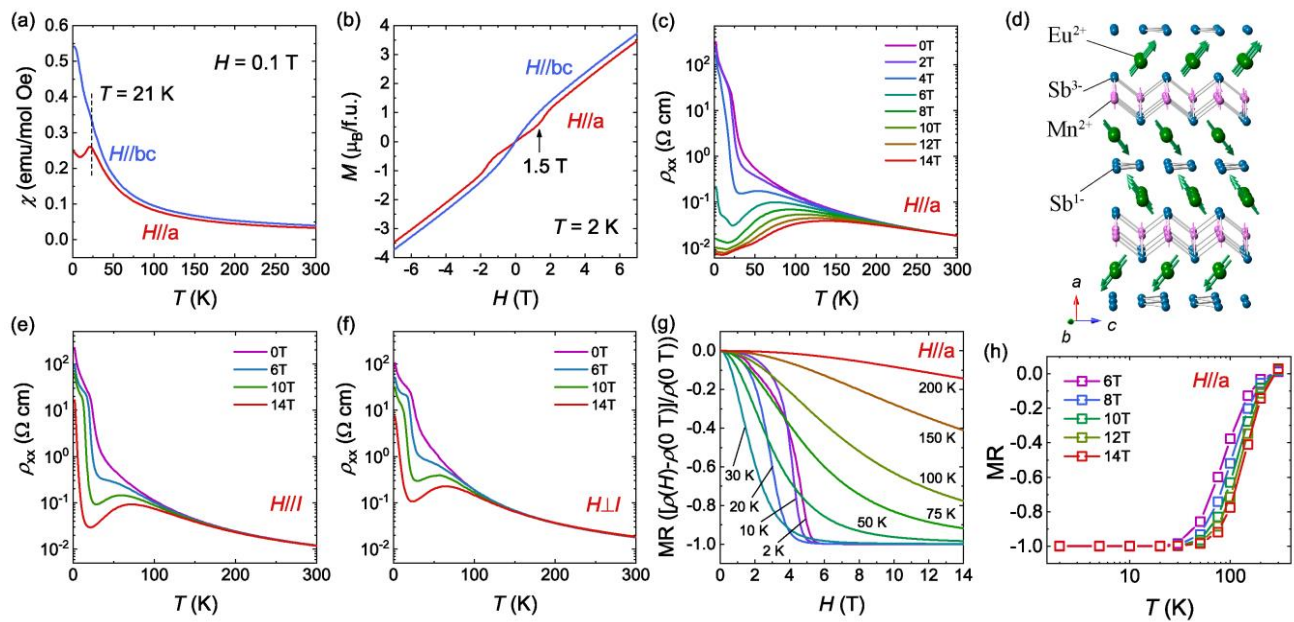

**Figure 2**

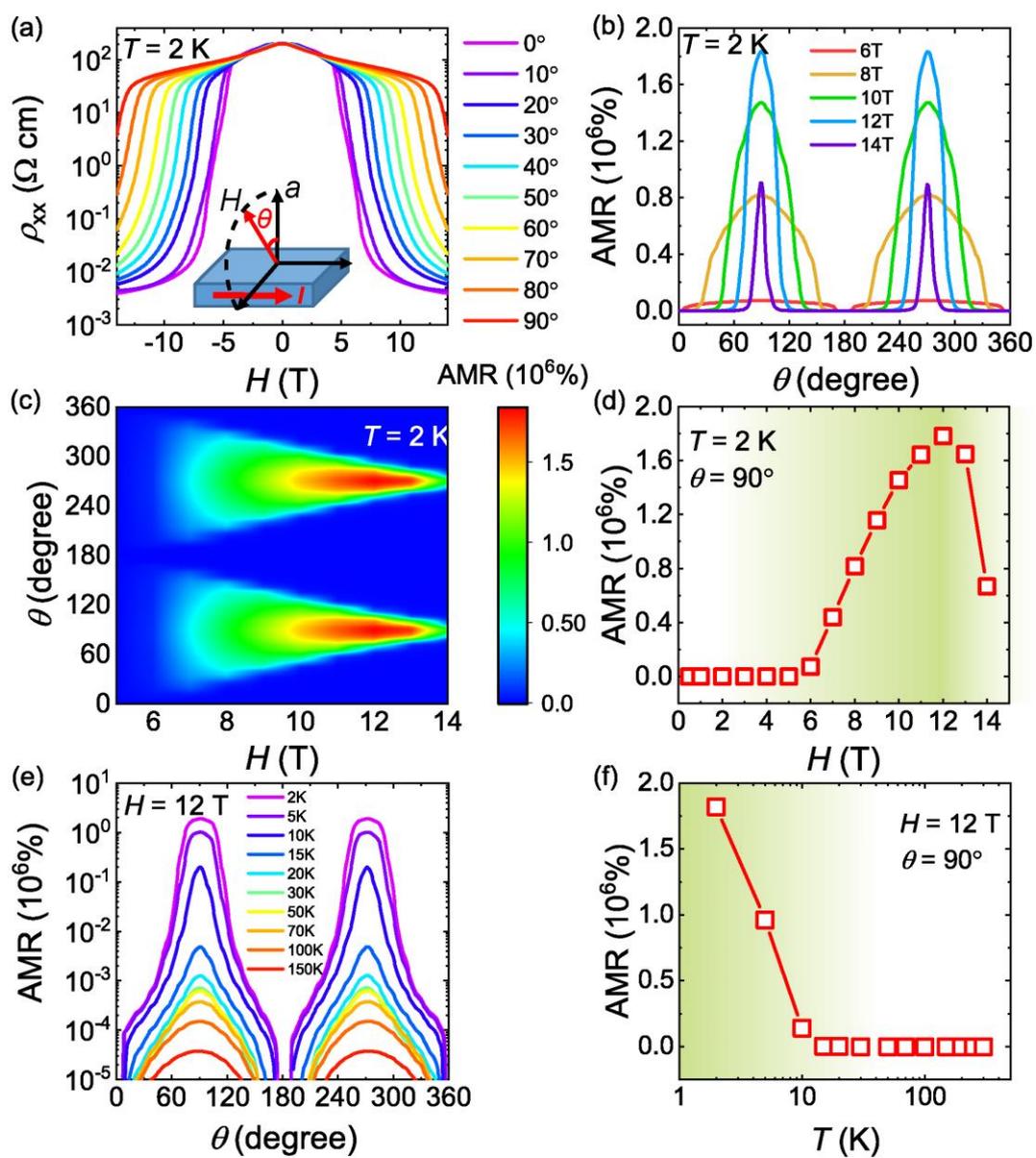

**Figure 3**

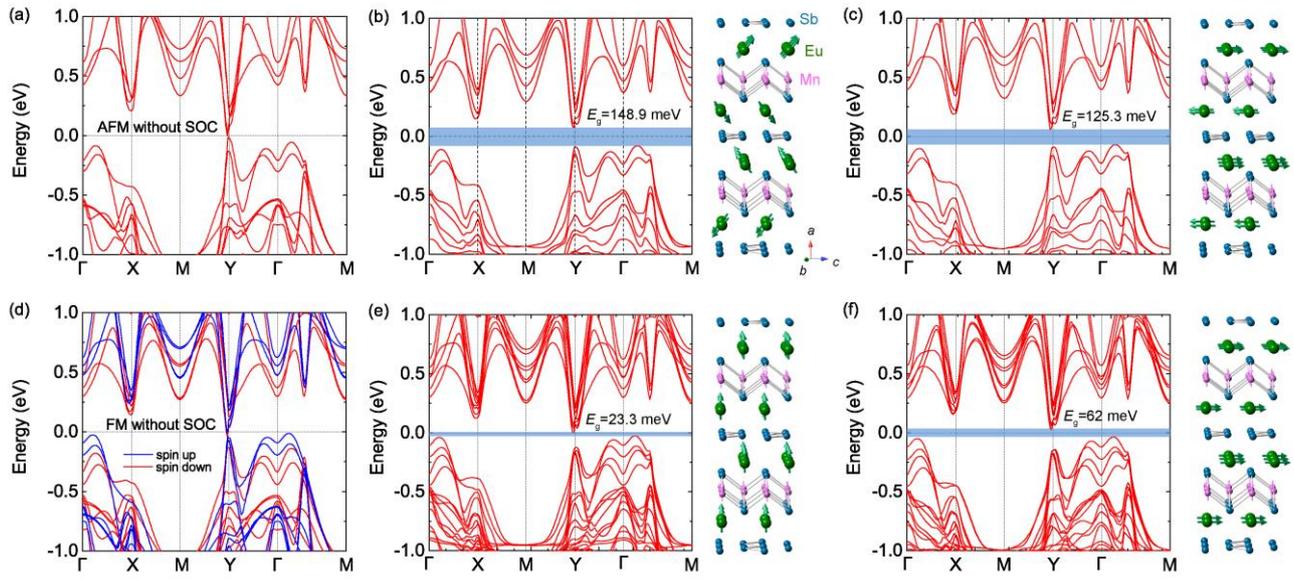

**Figure 4**

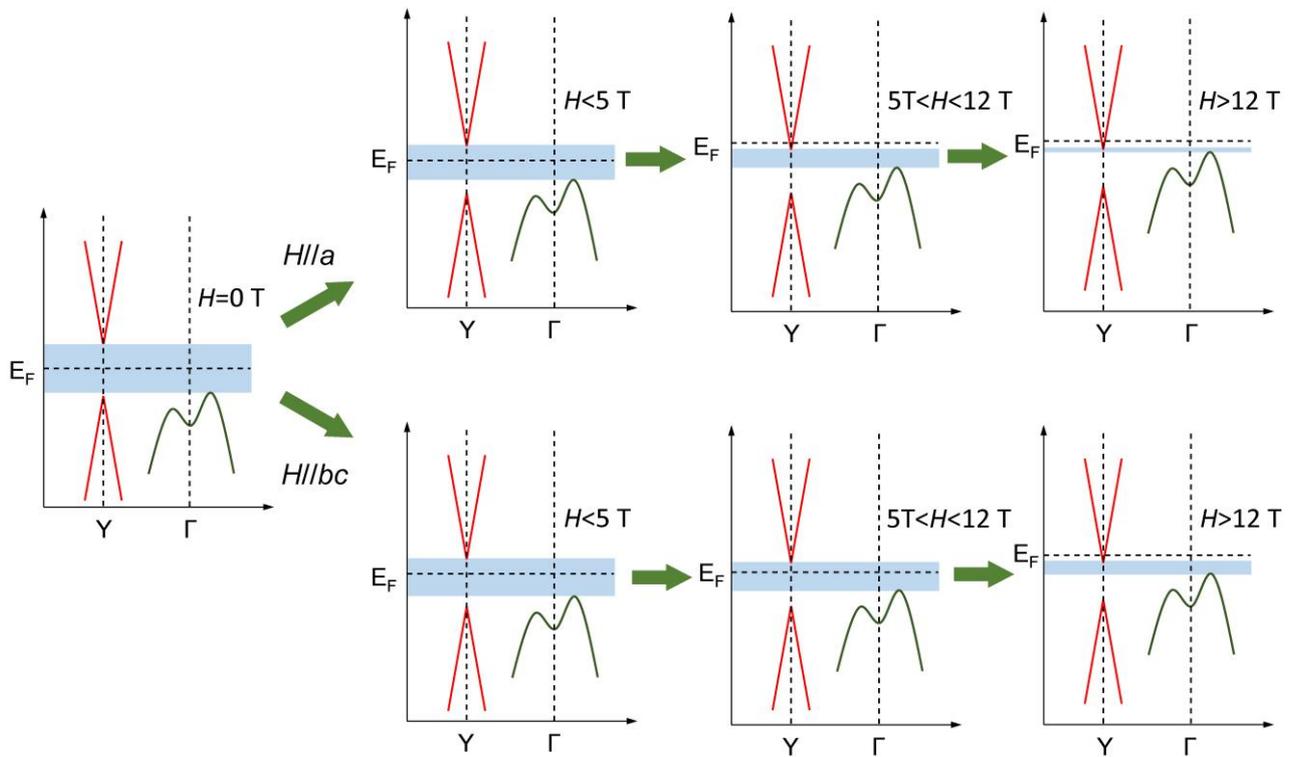